\documentclass{article}
\usepackage{amsfonts,epsfig}

\newtheorem{theorem}{Theorem}

\def\R{{\mathbb R}}
\def\C{{\mathbb C}}
\def\Z{{\mathbb Z}}

\def\D{{\cal D}}

\begin{document}

\title{On two-dimensional finite-gap potential Schr{\"o}dinger and Dirac
operators with singular spectral curves}
\author{Iskander A. TAIMANOV
\thanks{Institute of Mathematics, 630090 Novosibirsk, Russia;
taimanov@math.nsc.ru. The author was supported by the Russian Foundation for
Basic Researches, the Science Support Foundation and INTAS (grant no.
99-01782).}}
\date{}

\maketitle

\section{Introduction}

In the present paper we describe a wide class of two-dimensional
potential Schr{\"o}dinger and Dirac operators which are finite-gap
on the zero energy level and whose spectral curves at this level
are singular and, in particular, may have $n$-multiple points with
$n \geq 3$.

Dirac operators with such spectral curves are important for
the Weierstrass representation of tori in $\R^3$
\cite{T1,T2}. A study of finite-gap operators with
singular spectral curves which usually were not especially considered
because of the nongeneric situation is of a special interest
for differential geometry where singular curves may serve as
the spectral curves of smoothly immersed tori.
In particular, the spectral curves of tori in $\R^3$ obtained by a
rotation of circles lying in the plane
$y=0$ around the $x$ axis are rational curves with double points.

In the present paper
the problem of describing such operators is reduced to a problem
which involves only nonsingular curves by using the normalization of
spectral curves. That makes this description to be effective.

\section{Schr{\"o}dinger and Dirac operators which are fi\-ni\-te-gap on
the zero energy level}

\subsection{Two-dimensional operators which are finite-gap
the zero energy level}

The notion of a two-dimensional operator which is finite-gap at one
energy level was introduced by Dubrovin, Krichever, and Novikov
\cite{DKN} for the Schr{\"o}dinger operator.

First we recall the definition of a Floquet eigenfunction (or a
Bloch function) of a differential operator $L$ with periodic coefficients.
Let $L$ act on functions on $\R^n$ and let its coefficients be periodic
with respect to a lattice $\Lambda$ isomorphic to $\Z^n \subset \R^n$.
A solution of the equation
$$
L \psi = \lambda \psi, \ \ \ \lambda \in \C,
$$
is called a Floquet function (or a Bloch function) with the eigenvalue
$\lambda$, if for any vector $\gamma \in \Lambda$ we have
$$
\psi(x+\gamma) = e^{2\pi i \langle k,\gamma \rangle} \psi(x),
$$
where $\langle k,\gamma \rangle = \sum_{i=1}^n k_i \gamma_i$
is the standard scalar product.
The components of the vector $k=(k_1,\dots,k_n)$
are called the quasimomenta of $\psi$. We see that any Floquet function
defines a homomorphism
$$
\mu: \Lambda \to \C^\ast = \C \setminus\{0\}, \ \ \ \mu(\gamma) =
e^{2\pi i \langle k,\gamma \rangle}.
$$

By using the Keldysh theorem and assuming that the coefficients of
operators are bounded it has been proved that
for the Schr{\"o}dinger operator $\Delta + u$,
the heat operator $\partial_t - \Delta$ and
the two-dimensional Dirac operator,
the quasimomenta and the eigenvalues of
Floquet functions satisfy analytic relations (i.e. the dispersion laws
\cite{N}) and admissible tuples
$(k_1,\dots,k_n,\lambda)$ form an analytic subset $Q$ in
$\C^{n+1}$ \cite{Ku,T2}.
Here it is essential that these operators are hypoelliptic.
The set $Q$ is invariant under translations by vectors from the dual lattice
$\Lambda^\ast = \{\gamma^\ast : \langle \gamma^\ast , \gamma \rangle \in \Z
\ \mbox{for all} \ \gamma \in \Lambda\}$.
Therefore it is easier to consider the
quotient space  $Q/\Lambda^\ast$.

Let $n=2$, i.e. the operators are two-dimensional.
Then the intersection of $Q/ \Lambda^\ast$ with the plane
$\lambda=0$ is a complex curve (a Riemann surface) $\Gamma^\prime$,
on which Floquet functions are glued into a function
$\psi(x,P), P \in \Gamma^\prime$,
which is meromorphic on the surface outside finitely
many points. It is said that the operator $L$ is finite-gap at the zero
energy level  $\lambda=0$ if the curve $\Gamma$,
which is the normalization of the curve
$\Gamma^\prime$, is a curve of finite genus,
i.e. if $\Gamma$ is an algebraic curve.
This Riemann surface $\Gamma$ is called
the spectral curve of the operator $L$
at the zero energy level.

\subsection{The Schr{\"o}dinger operator}

The two-dimensional Schr{\"o}dinger operator with a magnetic field
has the form
\begin{equation}
\label{schrodinger}
L = \partial \bar{\partial} + A(z,\bar{z}) \bar{\partial} +
u(z,\bar{z}),
\end{equation}
where
$$
\partial = \frac{\partial}{\partial x} -
i \frac{\partial}{\partial y},\ \
\bar{\partial} = \frac{\partial}{\partial x} +
i \frac{\partial}{\partial y}, \ \
z = x+iy.
$$
Following \cite{DKN}, it is said that such an operator is finite-gap
at the zero energy level
\footnote{This definition as its analogue for Dirac operators (see \S 2.3)
is given only for operators with nonsingular spectral curves.
Operators with singular spectral curves appear from them in the
limit under a degeneration which could be rather complicated as we show
in this paper.}
if there exists

a) a nonsingular Riemann surface
of finite genus $g$ with two marked points
$\infty_\pm$ and local parameters
$k^{-1}_\pm$ near these points such that $k_\pm^{-1}(\infty_\pm)=0$;

b) an effective divisor (a formal sum of points on the surface)
$D = P_1 + \dots + P_g$ of degree $g$ and formed by points different from
$\infty_\pm$

\noindent
such that
there is a function $\psi=\psi(x,y,P)$
on $\Gamma$ meeting the following conditions

1) this function is meromorphic with respect to $P$
on $\Gamma \setminus\{\infty_\pm\}$, has poles only at
points from $D$ and the order of a pole is not greater than the number
of appearances of this  point in $D$: $(\psi) \geq -D$;

2) $\psi$ has the following asymptotics at $\infty_\pm$:
$$
\psi(x,y,P) \approx e^{k_+ z} (1 + \xi(x,y) k^{-1}_+ + O(k^{-2}_+))
\ \ \mbox{as} \ \ P \to \infty_+,
$$
$$
\psi(x,y,P) \approx c(x,y) e^{k_- \bar{z}} (1 + O(k^{-1}_-))
\ \ \mbox{as} \ \ P \to \infty_-;
$$

3) $\psi$ satisfies the equation $L\psi =0$
at every point $P \in \Gamma \setminus \{\infty_\pm\}$.

It follows from the theory of Baker--Akhieser functions that

1) for a generic divisor $D$
the data $(\Gamma,\infty_\pm,k_\pm,D)$
determines a unique function $\psi$ meeting conditions 1 and 2;

2) given a function $\psi$, a unique
Schr{\"o}dinger operator of the form (\ref{schrodinger})
can be constructed such that $L \psi =0$.
The explicit formulas take the form
$$
A = - \frac{\partial \log c}{\partial z}, \ \
u = - \frac{\partial \xi}{\partial \bar{z}}.
$$

As it is shown in \cite{VN,VN2}, if there exists a holomorphic involution of
$\Gamma$:
$$
\sigma: \Gamma \to \Gamma, \ \ \sigma^2 = 1,
$$
such that
$\sigma(\infty_\pm = \infty_\pm, \sigma(k_\pm) = - k_\pm$
and there exists a meromorphic differential (i.e. a $1$-form)
$\omega$ on $\Gamma$
with poles of the first order in the points
$\infty_+$ and $\infty_-$ and zeroes in the points from
$D + \sigma(D)$:
$$
D + \sigma(D) - \infty_+ - \infty_- \sim K(\Gamma)
$$
(the divisor in the left-hand side is equivalent to the canonical
divisor of the surface $\Gamma$),
then this operator is potential:
$c^2 \equiv 1$ and, therefore, $A = 0$.

If in addition there exists an antiholomorphic involution
$$
\tau: \Gamma \to \Gamma, \ \tau^2 = 1,
$$
such that
$$
\sigma\tau = \tau\sigma, \ \
\tau(D) = D,  \ \
\tau(\infty_\pm) = \infty_\mp, \ \
\tau(k_\pm) = \bar{k}_\mp,
$$
then the potential $u$ is real-valued.

It is easy to notice that the form
$\omega$ is invariant under $\sigma$ and hence it descends to a form
$\omega^\prime$ on the quotient surface
$\Gamma/\sigma$ and the form $\omega^\prime$
has $g$ zeroes and two simple poles.
Therefore the genus of $\Gamma/\sigma$ equals $g/2$ and the points
$\infty_\pm$ are exactly all fixed points of the involution $\sigma$
(notice that we assume that the surface $\Gamma$ is nonsingular).
The natural covering $\Gamma \to \Gamma_0 = \Gamma/\sigma$ is
two-sheeted and branched at the points $\infty_\pm$.

Generically the potential $u$ is quasi-periodic and if it is periodic then
the function $\psi(x,y,P)$ is
a Floquet function for every $P \in \Gamma \setminus \{\infty_\pm\}$
and the quasimomenta are locally holomorphic functions on the surface
$\Gamma$.

\subsection{The Dirac operator}

The Dirac operator (with potentials) has the form
$$
\D =
\left(
\begin{array}{cc}
0 & \partial \\
-\bar{\partial} & 0
\end{array}
\right) +
\left(
\begin{array}{cc}
U & 0 \\
0 & V
\end{array}
\right).
$$
It is said that it is finite-gap on the zero energy level
if there exists

a) a nonsingular Riemann surface $\Gamma$ of finite
genus $g$ with two marked points
$\infty_\pm$
and local parameters
$k^{-1}_\pm$ such that $k_\pm^{-1}(\infty_\pm)=0$;

b) an effective divisor $D = P_1 + \dots + P_{g+1}$
of degree $g+1$ formed by points which differ from
$\infty_\pm$,

\noindent
such that there is a vector function $\psi=(\psi_1,\psi_2)^\perp=
\psi(x,y,P)$
meeting the following conditions:

1) the function $\psi$ is meromorphic in $P$
on $\Gamma \setminus \{\infty_\pm\}$
and $(\psi) \geq -D$;

2) there are the following asymptotics:
$$
\psi(x,y,P) \approx e^{k_+ z}
\left[
\left(
\begin{array}{c}
1 \\ 0
\end{array}
\right) +
\left(
\begin{array}{c}
\xi^+_{1} \\ \xi^+_{2}
\end{array}
\right)
k^{-1}_+ + O(k^{-2}_+) \right]
\ \ \mbox{as} \ \ P \to \infty_+,
$$
$$
\psi(x,y,P) \approx e^{k_- \bar{z}}
\left[
\left(
\begin{array}{c}
0 \\ 1
\end{array}
\right) +
\left(
\begin{array}{c}
\xi^-_{1} \\ \xi^-_{2}
\end{array}
\right)
k^{-1}_- + O(k^{-2}_-) \right]
\ \ \mbox{as} \ \ P \to \infty_-;
$$

3) the equation $\D \psi = 0$ holds on $\Gamma \setminus \{\infty_\pm\}$.

As in the case of the Schr{\"o}dinger operator for a generic divisor $D$
the data $(\Gamma,\infty_\pm,k_\pm,D)$ determines a function satisfying
the conditions 1 and 2 uniquely and from this function one can construct
a unique operator $\D$ such that $\D \psi = 0$:
$$
U = -\xi^+_{2}, \ \ \ V = \xi^-_{1}.
$$
Again as in the case of the Schr{\"o}dinger operator
generically these potentials are quasi-periodic but when they are
periodic the functions $\psi(x,y,P)$ are Floquet functions whose
quasimomenta locally holomorphically depend on $P$.

If there exists a holomorphic involution
$\sigma: \Gamma \to \Gamma$ such that
$$
\sigma(\infty_\pm) = \infty_\pm, \ \ \sigma(k_\pm) = - k_\pm,
$$
and there exists a meromorphic differential $\omega$  with zeroes
in $D + \sigma(D)$ and two poles in the marked points
$\infty_\pm$
with the principal parts
$\pm k^2_\pm(1  + O(k^{-1}_\pm)) dk^{-1}_\pm$,
then the potentials $U$ and $V$ coincide: $U=V$ (\cite{T2}).

If there exists an antiholomorphic involution
$\tau: \Gamma \to \Gamma$,
such that
$$
\tau(\infty_\pm) = \infty_\mp, \ \ \tau(k_\pm) = - \bar{k}_\mp
$$
and there exists a meromorphic differential $\omega^\prime$  with zeroes in
$D + \tau(D)$ and two poles in $\infty_\pm$ with the principal parts
$k^2_\pm(1 + O(k^{-1}_\pm)) dk^{-1}_\pm$,
then the potentials $U$ and $V$ are real-valued:
$U=\bar{U}, V=\bar{V}$ (\cite{T2}).

For periodic operators these involutions are easily described on the
language of quasimomenta:
$$
\sigma(k_1,k_2) = (-k_1,-k_2), \ \
\tau(k_1,k_2) = (\bar{k}_1,\bar{k}_2),
$$
and the existence of them immediately follows from
from the spectral properties of the Dirac operator \cite{T3}.
\footnote{Notice that it needs to add the condition that the potentials
$U$ and $V$ are real-valued to the part 2 of Proposition 3 in \cite{T3},
where this condition is used in the proof.}
Obviously in this case these involutions commute.

{\sl Remark}. In papers \cite{T2,T3} we did some inaccuracy
assuming rather strong conditions for the differentials
$\omega$ and $\omega^\prime$ by demanding that they have
the following principal parts
$(\pm k^2_\pm + O(k^{-1}_\pm)) dk^{-1}_\pm$ and
$(k^2_\pm + O(k^{-1}_\pm)) dk^{-1}_\pm$ at points $\infty_\pm$.
But the exposed proofs work under weaker conditions mentioned above.
Indeed,

a) the differential $\psi_1(P)\psi_2(\sigma(P)) \omega$
has two poles of the first order at the points
$\infty_+$ and $\infty_-$ and the sum of residues equals
$-2\pi i(\xi_2^+ + \xi_1^-) =0$ which implies that $U=V$;

b) the differentials $\psi_1(P)\overline{\psi_1(\tau(P))} \omega^\prime$ and
$\psi_2(P)\overline{\psi_2(\tau(P))} \omega^\prime$
have first order poles at the points
$\infty_+$ and $\infty_-$ and the sums of residues
are equal to
$2\pi i(\xi^-_1 - \bar{\xi}^-_1)=0$ and $2\pi i(\xi^+_2 -
\bar{\xi}^+_2)=0$, respectively, which implies that $U=\bar{U}$ and
$V=\bar{V}$.

Let us demonstrate the involutions $\sigma$ and $\tau$
by the following simple example.

Let the potential $U=V=c$ equals a real nonzero constant $c$.
Then the spectral curve is the complex projective line $\Gamma =
\C P^1$ realized as the $\lambda$-plane completed by the infinity point
$\lambda = \infty$. There two marked points $\infty_\pm$ on $\Gamma$
such that $\lambda =\infty$ at $\infty_+$ and $\lambda =0$ at $\infty_-$.
Let us define near these points local parameters
$k_\pm^{-1}$ by the formulas
$$
k_+ = \lambda, \ \ \
k_- = -\frac{c^2}{\lambda}.
$$
The function $\psi$ takes the form
$$
\psi =
\left(
\begin{array}{c}
\frac{\lambda}{\lambda - c}
\exp\left(\lambda z - \frac{c^2}{\lambda} \bar{z} \right) \\
\frac{c}{c -\lambda}
\exp\left(\lambda z - \frac{c^2}{\lambda} \bar{z} \right)
\end{array}
\right),
$$
the divisor $D$ is just the point $\lambda=c$:
$$
D = c,
$$
and the involutions $\sigma$ and $\tau$ are defined by the formulas
$$
\sigma(\lambda) = - \lambda, \ \ \
\tau(\lambda) = \frac{c^2}{\bar{\lambda}}.
$$
The differentials $\omega$ and $\omega^\prime$ have the form
$$
\omega = \left(1-\frac{c^2}{\lambda^2}\right) d\lambda, \ \ \
\omega^\prime = \frac{(\lambda -c)^2}{\lambda^2} d\lambda.
$$

\section{Some facts on singular algebraic curves}

We expose some necessary facts on singular algebraic curves following
mostly to the book by Serre \cite{Serre}.

In the following we refer to
a (complex) algebraic curve as a curve.
Assuming that algebraic varieties are embedded into
$\C P^n$ we say that a mapping between them is regular if it is defined by
polynomials in homogeneous coordinates.

If a curve $\Gamma^\prime$ has singularities, then there is a
normalization
$$
\pi: \Gamma \to \Gamma^\prime,
$$
where

1) $\Gamma$ is a nonsingular curve with a finite set $S$ of
marked points on it and given an equivalence relation $\sim$ between
these points;

2) the mapping $\pi$ maps the set $S$ exactly onto the
singular locus $S^\prime$ of the curve $\Gamma^\prime$,
and the preimage of every point from
$S^\prime$ consists of a class of all equivalent points;

3) the mapping $\pi: \Gamma \setminus S \to \Gamma^\prime
\setminus S^\prime$
is a smooth one-to-one projection;

4) any regular mapping $F: X \to \Gamma^\prime$
of a nonsingular variety $X$ with an everywhere dense image $F(X) \subset
\Gamma^\prime$ descends through $\Gamma$:
$F = \pi G$ for some regular mapping $G: X \to \Gamma$.

We recall that for any point $P$ from an algebraic variety
there is a corresponding local ring ${\cal O}_P$ defined as ring of
functions on the variety which are induced by rational functions
$f/g$ where $f$ and $g$ are homogeneous
polynomials of the same degree and
$g(P) \neq 0$ (here we assume that the variety is embedded into
$\C P^n$). A point is nonsingular exactly
when its local ring is integrally closed.

For a point  $P \in \Gamma^\prime \setminus S^\prime$
its local ring ${\cal O}^\prime_P$ is ${\cal O}_{\pi^{-1}(P)} = {\cal O}_P$.
If $P \in S^\prime \subset \Gamma^\prime$, then
${\cal O}^\prime_P$ is a subring of the ring
$$
{\cal O}_P = \bigcap_{Q \to P}{\cal O}_Q
$$
and moreover ${\cal O}^\prime_P$ differs from ${\cal O}_P$ and for some
integer $n$ we have the following inclusions
$$
\C + R_P^n \subset {\cal O}^\prime_P \subset \C + R_P \subset
{\cal O}_P,
$$
where $R_P$ is an ideal, of the ring ${\cal O}_P$,
consisting of all functions vanishing at $\pi^{-1}(P)$.

There is a particular case of constructing a singular curve
$\Gamma_D$ from a nonsingular curve $\Gamma$ and an effective divisor
$D = \sum n_P P$ with degree
$\deg D = \sum n_P \geq 2$ on the curve $\Gamma$.
Let us denote by $S$
the set of points from $\Gamma$ with $n_P > 0$
(the support of the divisor) and put
$\Gamma_D = (\Gamma \setminus S) \cup \{\mbox{pt}\}$,
i.e. contract all points from $S$
into one point which we denote by $Q$.
Let us denote by $C_Q$ an ideal consisting of all functions $f$
which have at points $P \in S$ zeroes of order not less than $n_P$.
Now put ${\cal O}^\prime_Q = \C + C_Q$. This is the set consisting of
all functions which have the same value at all points $P \in S$
and whose first $(n_P-1)$ derivatives vanish at such a point.
The natural projection $\Gamma \to \Gamma_D$ is the normalization.
For $D = P_1 + \dots + P_n$ where all points $P_i$ are pair-wise
different the curve $\Gamma$ has an $n$-multiple point $Q$ with
different tangents.

To every singular point $P \in \Gamma^\prime$
there corresponds an integer-valued invariant
$$
\delta_P = \dim_\C {\cal O}_P/{\cal O}^\prime_P < \infty.
$$
It is obvious that for a singular point $Q$ of a curve $\Gamma_D$
we have
$$
\delta_Q = \dim {\cal O}_Q / (\C + C_Q) = \dim {\cal O}_Q/C_Q
-1 = \deg D -1.
$$

The genus of the nonsingular curve $\Gamma$ which is the normalized curve
is called the geometric genus of the curve
$\Gamma^\prime$ and it is denoted by $p_g(\Gamma^\prime)$,
and the quantity
$$
p_a(\Gamma^\prime) = p_g(\Gamma^\prime) + \sum_{P \in S} \delta_P
$$
is called the arithmetic genus of the curve $\Gamma^\prime$.

We notice that a meromorphic $1$-form (a differential)
$\omega$ on the curve $\Gamma$
is called a differential regular at a point $P \in \Gamma^\prime$,
if the equality
$$
\sum_{Q \to P} \mbox{Res}\ (f\omega) = 0
$$
holds for all $f \in {\cal O}^\prime_Q$. It is evident that there are more
regular differentials on
$\Gamma^\prime$ than regular differentials on
$\Gamma$, since regular differentials  on $\Gamma^\prime$ may have
have poles in the preimages of singular points.
For instance, for a curve
$\Gamma_D$ forms which are regular at a singular point $Q$
are distinguished by the following conditions:
a form $\omega$ may have poles only at
$P \in D$ with their orders not greater than
$n_P$ and
$$
\sum_{Q \to P} \mbox{Res}\ \omega = 0.
$$
It is easy to see that the dimension of the space of regular differentials
equals
$p_a(\Gamma^\prime)$.

Let the support of an effective divisor $D$ on the curve $\Gamma^\prime$
pairwise not intersect with the support of the divisor $S^\prime$.
We denote by $\dim L(D)$
the space of meromorphic functions on $\Gamma^\prime$ with poles only at
points from $D = \sum n_P P$ with orders not greater than
$n_P$,
and denote by $\Omega^\prime(D)$ the space of regular differentials
on $\Gamma^\prime$ which have at every point
$P \in S$ a zero whose order is not less than
$n_P$. The Riemann--Roch theorem reads
that
$$
\dim L(D) - \dim \Omega^\prime(D) =
\deg D + 1 - p_a(\Gamma^\prime).
$$
For a generic divisor $D$ we have
$\dim \Omega^\prime(D)=0$ and the Riemann--Roch theorem takes the form
$$
\dim L(D) =
\deg D + 1 - p_a(\Gamma^\prime).
$$

\section{Schr{\"o}dinger and Dirac operators corresponding to
singular spectral curves}

We consider curves of the form
$\Gamma_{B_1,\dots,B_n}$ which are successively constructed from
effective divisors $B_1,\dots,B_n$ and a curve $\Gamma$
by the same procedure which  constructs the curve
$\Gamma_D$ from $\Gamma$ and $D$.
Of course we assume that the supports of divisors
$B_i,i=1,\dots,n$, are pair-wise nonintersecting.

\begin{theorem}
1) Let $\Gamma^\prime = \Gamma_{B_1,\dots,B_n}$ be a singular curve,
let $\pi: \Gamma \to \Gamma^\prime$ be its normalization,
let $S_1,\dots,S_n$ be the supports of divisors
$B_1,\dots,B_n$ and let $Q_1,\dots,Q_n$
singular points of the curve $\Gamma^\prime$:
$\pi(S_i) = B_i, i=1,\dots,n$.

Let $\infty_+,\infty_-$ be a pair of different points from
$\Gamma^\prime \setminus \{Q_1,\dots,Q_n\}$
with local parameters $k^{-1}_{\pm}$ near these points such that
$k_\pm^{-1}(\infty_\pm) = 0$,
and let $D = P_1 + \dots + P_g$ be a generic effective divisor on
$\Gamma \setminus \{Q_1,\dots,Q_n,\infty_+,\infty_-\}$
of degree $g = p_a(\Gamma^\prime)$.

Then there exists an unique function
$\psi(x,y,P), P \in \Gamma^\prime$, such that
it is meromorphic everywhere outside the points $\infty_+$ and $\infty_-$
where it has the following asymptotics
$$
\psi(x,y,P) \approx e^{k_+ z} (1 + \xi(x,y) k^{-1}_+ + O(k^{-2}_+))
\ \ \mbox{as} \ \ P \to \infty_+,
$$
$$
\psi(x,y,P) \approx c(x,y) e^{k_- \bar{z}} (1 + O(k^{-1}_-))
\ \ \mbox{as} \ \ P \to \infty_-;
$$
and it has poles only in points from $D = P_1 + \dots + P_g = \sum n_P P$
of order not greater than $n_P$.

The function $\psi$ satisfies the equation $L\psi = 0$, where
the operator $L = \partial \bar{\partial} + A(z,\bar{z}) \bar{\partial} +
u(z,\bar{z})$ is uniquely reconstructed from $\psi$:
$$
A = - \frac{\partial \log c}{\partial z}, \
u = - \frac{\partial \xi}{\partial \bar{z}}.
$$

2) Let there be a holomorphic
involution $\sigma$ on $\Gamma^\prime$ which preserves the marked
points $\infty_\pm$ and inverts the local parameters near them:
$\sigma(\infty_\pm) = \infty_\pm$,
$\sigma(k^{-1}_\pm) = -k^{-1}_\pm$.
Let all singular points of the curve be fixed points of the involution
and let the involution preserve the branches of the curve in these points
(i.e. the pullback of the involution on $\Gamma$
preserves all points from $S$).

If there is a differential $\omega$ on $\Gamma_{2B_1,\dots,2B_n}$ such that
it is regular everywhere outside the points
$\infty_+$ and $\infty_-$ in which it has first order poles with
residues $\pm 1$ and
it has zeroes exactly in the points of $D + \sigma(D)$, then
$L$ is a potential operator:
$$
L = \partial \bar{\partial} + u.
$$

3) If in addition there is an antiholomorphic involution
$\tau: \Gamma^\prime \to \Gamma^\prime$ such that
$\sigma\tau = \tau\sigma, \tau(D) = D,
\tau(\infty_\pm) = \infty_\mp,
\tau(k_\pm) = \bar{k}_\mp$,
then the potential $u$ is a real-valued function.
\end{theorem}

{\sl Proof}. First we remark that for a nonsingular (smooth)
curve the statement
1 is the theorem by Dubrovin, Krichever, and Novikov
\cite{DKN}, and statements 2 and 3 were proved by Veselov and Novikov
\cite{VN}. We already exposed them in {\S} 2.

1) For a smooth curve of genus $g$ a construction of a function $\psi$ with
given properties was done in
\cite{DKN}. Let us apply this construction to the normalized curve
$\Gamma$ and the divisors $D_1 = P_1 + \dots + P_l,
D_2 = P_1 + \dots + P_{l-1} + P_{l+1}, \dots$, $D_{g-l+1} =
P_1 + \dots + P_{l-1} + P_g$, where $l = p_g(\Gamma^\prime)$
is the genus of the curve
$\Gamma$. We obtain the functions
$\psi_1,\dots,\psi_{g-l+1}$. The desired function
$\psi$ has the form
$$
\psi = c_1 \psi_1 + \dots + c_{g-l+1} \psi_{g-l+1},
$$
where the coefficients $c_i$ are found by using two conditions:

1) the function $\psi$ descends to a function on $\Gamma^\prime$ (these are
$(g-l) =
(p_a(\Gamma^\prime) - p_g(\Gamma^\prime))$ equations);

2) the asymptotics $e^{k_+ z}$ holds at $\infty_+$.

For instance, given $B = Q_1 + \dots + Q_{g-l+1}$,
the first condition is written as
$$
\psi(Q_1) = \psi(Q_2), \ \psi(Q_1)=\psi(Q_3), \dots, \
\psi(Q_1) = \psi(Q_{g-l+1}),
$$
and for $B = mQ$ it takes the form
$$
\frac{\partial \psi(Q)}{\partial w} = \dots =
\frac{\partial^{m-1} \psi(Q)}{\partial^{m-1} w}
=0,
$$
where $w$ is a local parameter on $\Gamma$ near $Q$.
The second condition looks the same in both cases:
$$
c_1 + \dots + c_{g-l+1} = 1.
$$

The uniqueness of the functions $\psi_i$ follows from the
Riemann--Roch theorem
\cite{DKN}. Together with conditions for $c_i$ this implies
the uniqueness of
the function $\psi$ for a generic divisor $D$.
For given functions $A$ and $u$ the function
$L\psi$ is proportional to $\psi$, but its asymptotics as
$P \to \infty_+$ are $\alpha(z,\bar{z}) e^{k_+z}k^{-1}_+$,
and therefore $L\psi$ vanishes everywhere \cite{DKN}.

2) Let us consider the form $\psi(P)\psi(\sigma(P))\omega(P)$.
It is meromorphic and the points $\infty_+$ and $\infty_-$ have residues
$1$ and $-c^2$, respectively. Let
$Q_i$ be a singular point, of $\Gamma^\prime$, corresponding to the divisor
$B_i = \sum n_Q Q$.
Since the function $\psi(P)\psi(\sigma(P))$ is invariant under
the involution,
near a fixed point $Q$ this function expands in a series in even
degrees of a local parameter $k$ where
$k(Q)=0, \sigma(k)=-k$:
$$
\psi(P)\psi(\sigma(P)) = \psi(Q)^2 + a_1 k^2 + \dots + a_n k^{2n}
+ \dots, \ \ k(P) = k,
$$
and moreover, since $\psi$ descends to a function on
$\Gamma^\prime$, we have
$a_j = 0$ for $j < n_P$.
Since the form $\omega$ is regular on $\Gamma_{2B_1,\dots,2B_n}$,
it has a pole of order not greater than $2n_P$ at the point $P$:
$$
\omega = b_{2n_P} \frac{dk}{k^{2n_P}} + \dots + b_1 \frac{dk}{k} +
\, \mbox{the regular terms}.
$$
Therefore, there is the following formula for the residue:
$$
\mbox{Res}\,[\psi(P)\psi(\sigma(P))\omega]\big|_{P=Q}
= b_1(Q)\psi(Q_i)^2.
$$
The regularity of $\omega$ at
$Q_i \in \Gamma^\prime$
implies that the sum of the residues of $\omega$ over
the preimage of this point vanishes:
$$
\psi(Q_i)^2 \sum_{Q \in S_i}  b_1(Q) = 0.
$$
Therefore every singular point
$Q_i$ does not contribute to the sum of the residues
of the differential $\psi(P)\psi(\sigma(P))\omega$ and,
since this sum equals
$$
1 - c^2 = 0,
$$
we have $c^2 =0$ and $A = - \frac{\partial \log c}{\partial z} = 0$.

3) For nonsingular curves this statement was proved by Novikov and Veselov
and in the nonsingular case the proof works without changes as follows.
Let us consider the expansion for the function $\psi$ at
$\infty_-$:
$$
\psi \sim ce^{k_-\bar{z}}(1 + \eta k^{-1}_- + O(k^{-2}_-)).
$$
Since $L = \partial \bar{\partial} + u$ and $c^2 =1$, we obtain
$$
L\psi = ce^{k_-\bar{z}}( (u+\partial \eta) + O(k^{-1}_-)) = 0,
$$
which implies the formula for the potential in terms of the asymptotics
of $\psi$ near $\infty_-$:
$$
u = -\partial \eta.
$$
Since the function $\psi$ is uniquely reconstructed from the data
$\Gamma^\prime,\infty_{\pm},k_{\pm},D$, this uniqueness
theorem implies the equality
$$
\overline{\psi(\tau(P))} = \bar{c} \psi(P) = c\psi(P).
$$
In particular, $\xi = \bar{\eta}$, and comparing the two
formulas for $u$ we have
$$
u = - \bar{\partial} \xi = -\partial \eta,
$$
We conclude that the potential $u$ is real-valued: $u = \bar{u}$.

This proves the theorem.

The essential part of this theorem is the condition for the differential
$\omega$ which has to be regular not on
$\Gamma_{B_1,\dots,B_n}$ but on $\Gamma_{2B_1,\dots,2B_n}$.
Together with that the proof of the following theorem on Dirac operators
is obtained  by a modification of the proof for a nonsingular case
(see \cite{T2}).
\footnote{Although it is not difficult to formulate the following theorem
using Theorem 1 and the results exposed in \S 2.3
we do that for the completness of
exposition.}

\begin{theorem}
1) Let $\Gamma^\prime = \Gamma_{B_1,\dots,B_n}$ be a singular curve,
let $\pi: \Gamma \to \Gamma^\prime$ be its normalization,
let $S_1,\dots,S_n$ be the supports of the divisors
$B_1,\dots,B_n$ and let $Q_1,\dots,Q_n$ be singular points of
$\Gamma^\prime$: $\pi(S_i) = B_i, i=1,\dots,n$.

Let $\infty_+,\infty_-$ be a pair of different points from
$\Gamma^\prime \setminus \{Q_1,\dots,Q_n\}$
with local parameters $k^{-1}_{\pm}$ near these points such that
$k_\pm^{-1}(\infty_\pm) = 0$,
and let $D = P_1 + \dots + P_g +P_{g+1}$ be a generic effective divisor on
$\Gamma \setminus \{Q_1,\dots,Q_n,\infty_+,\infty_-\}$
of degree $g+1 = p_a(\Gamma^\prime)+1$.

Then there exists a unique vector function
$\psi(x,y,P), P \in \Gamma^\prime$, such that
it is meromorphic everywhere outside the points $\infty_+$ and $\infty_-$
where it has the following asymptotics
$$
\psi(x,y,P) \approx e^{k_+ z}
\left[
\left(
\begin{array}{c}
1 \\ 0
\end{array}
\right) +
\left(
\begin{array}{c}
\xi^+_{1} \\ \xi^+_{2}
\end{array}
\right)
k^{-1}_+ + O(k^{-2}_+) \right]
\ \ \mbox{as} \ \ P \to \infty_+,
$$
$$
\psi(x,y,P) \approx e^{k_- \bar{z}}
\left[
\left(
\begin{array}{c}
0 \\ 1
\end{array}
\right) +
\left(
\begin{array}{c}
\xi^-_{1} \\ \xi^-_{2}
\end{array}
\right)
k^{-1}_- + O(k^{-2}_-) \right]
\ \ \mbox{as} \ \ P \to \infty_-;
$$
and it has poles only in points from
$D = P_1 + \dots + P_{g+1} = \sum n_P P$
of order not greater than $n_P$.

The function $\psi$ satisfies the equation $\D \psi = 0$,
where the operator
$$
\D =
\left(
\begin{array}{cc}
0 & \partial \\
-\bar{\partial} & 0
\end{array}
\right) +
\left(
\begin{array}{cc}
U & 0 \\
0 & V
\end{array}
\right)
$$
is uniquely reconstructed from $\psi$:
$$
U = - \xi^+_2, \ \ V = \xi^-_1.
$$

2) Let there be a holomorphic
involution $\sigma$ on $\Gamma^\prime$ which preserves the marked
points $\infty_\pm$ and inverts the local parameters near them:
$\sigma(\infty_\pm) = \infty_\pm$,
$\sigma(k^{-1}_\pm) = -k^{-1}_\pm$.
Let all singular points of the curve be fixed points of the involution
and the involution preserves the branches of the curve at these points
(i.e. the pullback of the involution on $\Gamma$
preserves all points from $S$).

If there is a differential $\omega$ on $\Gamma_{2B_1,\dots,2B_n}$ such that
it is regular everywhere outside the points
$\infty_+$ and $\infty_-$ in which it has second order
poles with the principal parts
$\pm k^2_\pm(1 + O(k^{-1}_\pm))dk^{-1}_\pm$ and its has
zeroes exactly at points from
$D + \sigma(D)$, then the potentials $U$ and $V$ coincide:
$$
U=V.
$$

3) Let there be an antiholomorphic involution
$\tau: \Gamma^\prime \to \Gamma^\prime$
such that it interchanges the points $\infty_+$ and $\infty_-$:
$$
\tau(\infty_\pm) = \infty_\mp, \ \ \tau(k_\pm) = -\bar{k}_\mp,
$$
and its pullback onto $\Gamma$ preserves all points from
$S_1 \cup \dots \cup S_n$,
changing local parameters $k$ by the formula
$\tau(k) = -\bar{k}$.

Let there exist a differential $\omega^\prime$, on
$\Gamma_{2B_1,\dots,2B_n}$, which is regular everywhere outside the points
$\infty_+$ and $\infty_-$ in which it has second order
poles with the principal parts
$k^2_\pm(1 + O(k^{-1}_\pm))dk^{-1}_\pm$ and which has zeroes
exactly at points from
$D + \tau(D)$. Then the potentials
$U$ and $V$ are real-valued:
$$
U = \bar{U}, \ \ \ V = \bar{V}.
$$
\end{theorem}

{\sl Remarks.}

\begin{figure}[ht]
\begin{center}
\epsfig{file=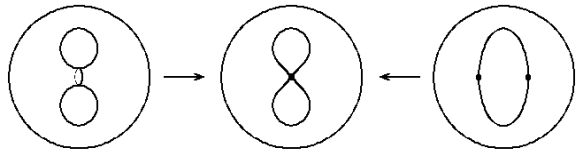,height=20mm,clip=}
\caption{A double point with preserved branches.}
\label{pic1}
\end{center}
\end{figure}

1) Potential Schr{\"o}dinger operators whose spectral curves
have only double points were described in the initial paper
by Novikov and Veselov
\cite{VN} as obtained by contracting invariant
cycles on nonsingular curves into points
(see pic. \ref{pic1} where it is demonstrated by a deformation of
an elliptic curve. Here by an involution we mean a rotation by
$\pi$ around the horizontal line which lies in the plane of the picture
and the right arrow denotes the normalization mapping).

These potentials are described in terms of the Prym theta functions of
double-sheeted coverings of singular curves.
The curve $\Gamma$ has double points which are fixed points of
the involution and moreover the branches are not permuted by the involution
(i.e. on the normalized curve the preimages of such fixed points are fixed
by the involution, see pic. \ref{pic1}).
In this case a complete Prym variety is defined as a principally-polarized
Abelian variety in a limit under the degeneration of nonsingular curves
$\Gamma$. In this limit the Jacobian variety of
the curve $\Gamma/ \sigma$ is a non-complete
Abelian variety, i.e. it has a form $\C^g / Z$ where the rank of
a lattice $Z$ is less than $2g$.

\begin{figure}[ht]
\begin{center}
\epsfig{file=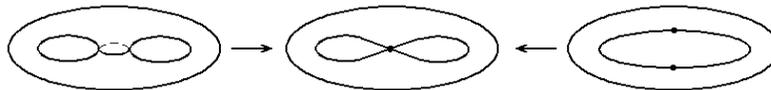,height=12mm,clip=}
\caption{A double point with permuted branches.}
\label{pic2}
\end{center}
\end{figure}

If the branches in a double point
are permuted by the involution (see pic. \ref{pic2}), then the
limiting Prym variety is not complete but the limiting Jacobian variety
of the curve $\Gamma/ \sigma$ is complete.

2) As we already pointed out above
(in {\S} 2), for a nonsingular curve corresponding
to a finite-gap potential Schr{\"o}dinger operator the following
equality holds
$$
p_a(\Gamma) = 2p_a(\Gamma/\sigma),
$$
which relates the arithmetic genera of the curve and its quotient
under the involution $\sigma$.
It also holds for curves with double points.
Let $\Gamma_B$ be a curve such that
$\Gamma$ is a hyperelliptic curve of genus two,
$\sigma$ is the hyperelliptic involution, and
$\infty_\pm \cup B$ are six fixed points of $\sigma$:
($B = Q_1 + \dots + Q_4$). We have $p_a(\Gamma_B) =
p_g(\Gamma)+3 = 5$ and $\Gamma_B/\sigma$ is a rational curve (a sphere)
with a quadruple point. Therefore, for this example we have
$$
p_a(\Gamma^\prime) = 5, \ \ p_a(\Gamma^\prime/\sigma) = 3.
$$
In this case the potential of the Schr{\"o}dinger
operator is written in terms of
the Prym variety of the double-sheeted covering
$\Gamma^\prime \to \Gamma^\prime/\sigma$.
This Prym variety is isomorphic to the Jacobian variety of the curve
$\Gamma$.

3) Notice, that, by the Krichever theorem \cite{K2},
all smooth real potentials of the Schr{\"o}dinger operator are
approximated by the potentials which are finite-gap on the zero energy level
to arbitrary precision. Moreover this approximation is generated by
approximations of their Floquet spectra by nonsingular spectral curves of
finite genera.
For the Dirac operator the analogue of this theorem is not proved
but it is clear that its proof can be obtained by some
modification of Krichever's reasonings.

4) One-dimensional Schr{\"o}dinger operators $L = \partial^2_x + u$
have hyperelliptic spectral curves which parameterize Bloch functions
for all values of the energy \cite{DMN}. Degenerations of such curves
(inside the class of hyperelliptic curves) lead to
soliton potentials on the background of finite-gap potentials.
\cite{K3}.

For instance, on the language of {\S} 3
the rational potential $u(x) = 2x^{-2}$
is constructed from the curve $\Gamma: w^2 = E$ and the point
$P$ with $E=w=0$. Its spectral curve is $\Gamma_{2P}$, and the spectral curve
of the potential $u(x) = 2/\cosh^2(x)$ has the form
$\Gamma_{Q+\sigma(Q)}$, where $\sigma$ is the hyperelliptic involution and
$Q \neq \sigma(Q)$.

5) Notice that the operators $L = i\partial_y - \partial^2_x + u$ whose
spectral curves on the zero energy level
are singular and normalized by a rational curve
were described in the paper \cite{DKMM}.


\begin{thebibliography}{MMM}

\bibitem{T1}
Taimanov, I. A.
Modified Novikov--Veselov equation and
differential geometry of surfaces.
Translations of the Amer.
Math. Soc., Ser. 2. 1997. V. 179. P. 133--151.

\bibitem{T2}
Taimanov, I.A.
The Weierstrass representation of closed surfaces in ${\mathbb R}^3$.
Functional Anal. Appl. {\bf 32}:4 (1998), 49--62.

\bibitem{DKN}
Dubrovin, B.A., Krichever, I.M., and Novikov, S.P.
The Schr{\"o}dinger equation in a periodic field and Riemann surfaces,
Soviet Math. Dokl. {\bf 17} (1976), 947--952.

\bibitem{N}
Novikov, S.P.
Two-dimensional Schr{\"o}dinger operators in periodic fields,
Journal of Soviet Mathematics {\bf 28} (1985), 1--20.

\bibitem{Ku}
Kuchment, P.
Floquet theory for partial differential equations.
Birk\-h{\"a}user, Basel, 1993.

\bibitem{VN}
Veselov, A.P., and Novikov, S.P.
Finite-zone two-dimensional Schr{\"o}\-din\-ger operators. Potential
operators,
Soviet Math. Dokl. {\bf 30} (1984), 705--708.

\bibitem{VN2}
Veselov, A.P., and Novikov, S.P.
Finite-zone two-dimensional potential Schr{\"o}dinger operators.
Explicit formulas and evolution equations.
Soviet Math. Dokl. {\bf 30} (1984), 588--591.

\bibitem{T3}
Taimanov, I.A.
Finite-gap solutions of the modified Novikov--Veselov equation,
their spectral properties, and applications.
Siberian Math. Journal {\bf 40} (1999), 1146--1156.

\bibitem{Serre}
Serre, J.-P.
Algebraic groups and class fields.
Graduate Texts in Mathematics, {\bf 117}.
Springer-Verlag, New York, 1988.

\bibitem{K2}
Krichever, I.M.
Spectral theory of two-dimensional periodic Schr{\"o}\-din\-ger operators
and its applications,
Russian Math. Surveys {\bf 44}:2 (1989), 145--225.

\bibitem{DMN}
Dubrovin, B.A., Matveev, V.B., and Novikov, S.P.
Non-linear equations of Korteweg--de Vries type, finite-zone linear
operators, and Abelian varieties,
Russian Math. Surveys {\bf 31}:1, 59--146.

\bibitem{K3}
Krichever, I.M.,
Potentials with zero coefficient of reflection
against a background of finite-zone potentials.
Functional Anal. Appl. {\bf 9}:2 (1975), 161--163.


\bibitem{DKMM}
Dubrovin, B.A., Malanyuk, T.M., Krichever, I.M., and
Makhankov, V.G.
Exact solutions of the time-dependent Schr{\"o}dinger equation
with self-consistent potentials.
Soviet J. Particles and Nuclei {\bf 19}:3 (1988), 252--269.

\end{thebibliography}
\end{document}